\documentclass[twocolumn,showpacs,preprintnumbers,amsmath,amssymb]{revtex4}

\usepackage{graphicx}
\usepackage{dcolumn}
\usepackage{bm}

\begin{document}

\preprint{APS/PRB}

\title{Static magnetic order in Na$_{0.75}$CoO$_2$ detected by muon spin rotation and relaxation}

\author{Jun Sugiyama}
 \email{sugiyama@iclab.tytlabs.co.jp}
\author{Hiroshi Itahara}%
\affiliation{%
Toyota Central Research and Development Labs., Inc., Nagakute, Aichi 480-1192, Japan}%

\author{Jess H. Brewer}
\affiliation{
TRIUMF, Canadian Institute for Advanced Research and
Department of Physics and Astronomy, University of British Columbia,
Vancouver, BC, V6T 1Z1 Canada
}%

\author{Eduardo J. Ansaldo}
\affiliation{
Department of Physics, University of Saskatchewan,
Saskatoon SK, S7N 5A5 Canada
}%

\author{Teruki Motohashi}
\author{Maarit Karppinen}
\author{Hisao Yamauchi}
\affiliation{
Materials and Structures Laboratory, Tokyo Institute of Technology, 
Nagatsuta, Midori-ku, Yokohama 226-8503, Japan
}%

\date{\today}

\begin{abstract}
The nature of the magnetic transition of the
Na-rich thermoelectric Na$_{0.75}$CoO$_2$ at 22K was studied
by positive muon-spin-rotation and relaxation ($\mu^+$SR)
spectroscopy, using a polycrystalline sample in the
temperature range between 300 
and 2.5~K. Zero field $\mu$SR measurements indicated
the existence of a 
static internal magnetic field at temperatures below
22~K (= $T_{\rm m}$).
The observed muon spin precession signal below $T_{\rm
m}$ consisted of 
three components with different precession
frequencies, corresponding to three inequivalent
muon$^+$ sites in the Na$_{0.75}$CoO$_2$ 
lattice. The total volume fraction of the three
components was estimated as 
$\sim$21\% at 2.5~K; thus, this magnetic transition
was not induced by 
impurities but is an intrinsic change in the magnetism
of the sample, although 
the sample was magnetically inhomogeneous otherwise.
On the other hand, a similar 
experiment on a Na$_{0.65}$CoO$_2$ sample exhibited no
magnetic 
transition down to 2.5~K; which indicates that the
average valence of  the Co 
ions is responsible for inducing the
magnetic transition at 22~K. 

\end{abstract}

\pacs{76.75.+i, 75.30.-m, 72.15.Jf}%
\keywords{Thermoelectric layered cobaltites, magnetism,
  muon spin rotation}
\maketitle

\section{\label{sec:level1}Introduction}

The layered cobaltite, Na$_x$CoO$_2$ with $x \sim$
0.5,\cite{1,2,3} 
is known to exhibit metallic conductivity $\sigma$ and
an extraordinarily large Seebeck coefficient $S$
(above +100~$\mu$V/K at 300~K) simultaneously,
probably due to a strong correlation between the 3$d$
electrons of the Co ions.\cite{4} 
The crystal structure of Na$_x$CoO$_2$ with 0.5 $\leq x \leq$
0.75 was reported to be a bronze-type
hexagonal system of space group $P6_3/mmc$ 
($a=0.2833$~nm and $c=1.082$~nm for $x$ =
0.71).\cite{5} 
In this structure, the single CoO$_2$ sheets and the
single disordered Na planes 
form alternating stacks along the hexagonal $c$ axis. 

The CoO$_2$ sheets, in which a
two-dimensional-triangular lattice of Co 
ions is formed by a network of edge-sharing CoO$_6$
octahedra, are believed 
to be the conduction planes. This is because the
CoO$_2$ sheet is a common 
structural component for all known thermoelectric
layered cobaltites, {\sl 
i.e.}, Na$_x$CoO$_2$, Ca$_3$Co$_4$O$_9$\cite{6,7,8}
and 
Bi$_2$Sr$_2$Co$_2$O$_y$.\cite{9,10,11} Moreover, if
the interaction between 3$d$ 
electrons plays a significant role on their transport
properties, such interaction should also affect the
magnetism of these cobaltites. 

Recently, Motohashi {\sl et al.} studied the bulk
susceptibility of polycrystalline 
Na$_{0.75}$CoO$_2$ and reported the existence of a
magnetic transition at 22~K (= $T_{\rm m}$)
accompanying the increase in the slope of the
resistivity-{\sl vs.}-$T$ curve and the appearance of
the large positive magnetoresistance effect. No 
transitions were found in Na$_{0.65}$CoO$_2$ down to
2~K.\cite{12} Interestingly, 
both $\sigma$ and $S$ of Na$_{0.75}$CoO$_2$ were
significantly larger than those of 
Na$_{0.65}$CoO$_2$.\cite{13} In other words, the
thermoelectric properties 
of Na$_x$CoO$_2$ seem to be enhanced by the magnetic
interaction between 
3$d$ electrons which induces the magnetic
transition. 

The measurements on heat capacity ($C_{\rm p}$) and spontaneous 
magnetization suggested that only a very small
fraction (less than 1\%) of the 
Na$_{0.75}$CoO$_2$ sample changed to the magnetic phase
even at 
2~K,\cite{12} although the sample was confirmed to be
single phase by 
powder X-ray diffraction analysis at ambient
temperature, leaving open the 
possibility that the magnetic transition is due
to an undetected impurity 
phase. However, such small impurity phase is unlikely
to induce the observed change 
in the transport properties below $T_{\rm m}$.
Therefore, to investigate the magnetism of 
Na$_x$CoO$_2$ in greater detail, 
we have measured both weak ($\sim$ 100~Oe)
transverse-field 
positive muon spin rotation and relaxation
(wTF-$\mu^+$SR) 
and zero field (ZF-) $\mu^+$SR spectra 
in both Na$_{0.75}$CoO$_2$ and Na$_{0.65}$CoO$_2$ at
temperatures below 300~K. 
The former method is sensitive to local magnetic order
{\it via\/} the shift of the $\mu^+$ spin precession
frequency 
and the enhanced $\mu^+$ spin relaxation, 
while ZF-$\mu^+$SR is sensitive to weak local magnetic
[dis]order 
in samples exhibiting quasi-static paramagnetic
moments.

\section{\label{sec:level1}Experimental}

Samples of Na$_{0.75}$CoO$_2$ and Na$_{0.65}$CoO$_2$
were synthesized by a 
modified solid state reaction technique, {\sl i.e.}, a
"rapid heat-up" 
technique, which was developed by Motohashi {\sl et al.}\cite{13} to control the Na content precisely, using reagent-grade Na$_2$CO$_3$ and
Co$_3$O$_4$ powders as 
starting materials. The mixed powder was placed into
the furnace, which was 
preheated at 750$^{\rm o}$C, and fired for 12 hours.
The fired powder was thoroughly ground 
and pressed into a plate of 10~mm length, 
15~mm width and 3~mm 
thickness, and then the plate was sintered at
900$^{\rm o}$C for 12 hours. 

Powder X-ray diffraction studies indicated that the
samples were single 
phase of a hexagonal structure of space group
$P6_3/mmc$, {\sl i.e.}, $\gamma$-Na$_x$CoO$_2$ phase. 
The lattice 
parameters of the Na$_{0.75}$CoO$_2$ sample were
calculated as $a$ = 
0.2828~nm and $c$ = 1.0884~nm, and for
Na$_{0.65}$CoO$_2$, $a$ = 0.2826~nm and 
$c$ = 1.0926~nm. The preparation and characterization
of the samples were 
reported in detail elsewhere.\cite{12,13} 
The $\mu^+$SR experiments were performed on the {\sf
M20} surface muon beam 
line at TRIUMF. The experimental setup and techniques
were described elsewhere.\cite{14} 

\section{\label{sec:level1}Results}

The wTF-$\mu^+$SR spectra for both samples were fitted
in the time domain 
with an exponentially damped  (relaxing) precessing
signal: 
\begin{eqnarray} 
  A_0 \, P(t) &=& A_{\rm TF} \, e^{- \lambda_{\rm TF}t} \, 
  \cos (\omega_\mu t + \phi) \; , 
\label{eq:TFfit} 
\end{eqnarray} 
where $A_0$ is the initial asymmetry, 
$P(t)$ is the muon spin polarization function, 
$\omega_\mu$ is the muon Larmor frequency, 
$\phi$ is the initial phase of the precession and 
$A_{\rm TF}$ and $\lambda_{\rm TF}$ are the asymmetry 
and exponential relaxation rate. 
\begin{figure} 
\includegraphics[width=8cm]{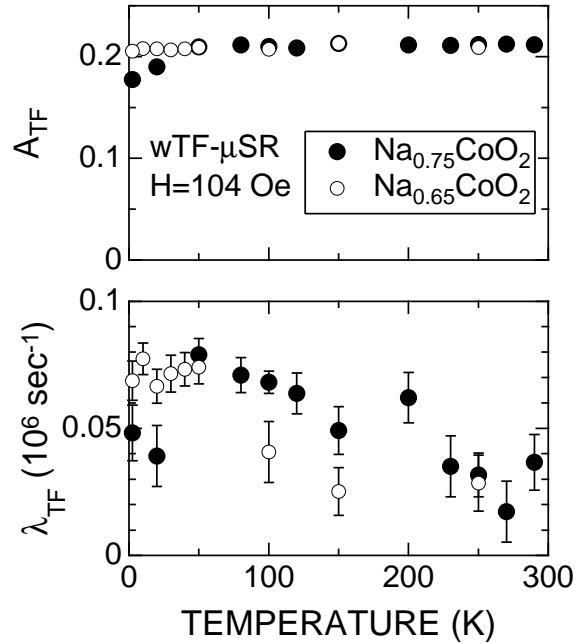}
\caption{\label{fig:wTF-muSR} The temperature
dependences of (a) $A_{\rm 
TF}$ and (b) $\lambda_{\rm TF}$ for Na$_{0.75}$CoO$_2$
and 
Na$_{0.65}$CoO$_2$. The data were obtained from fits
of the wTF-$\mu^+$SR 
time spectra using eq. (1).} 
\end{figure} 

Figures~\ref{fig:wTF-muSR}(a) and
\ref{fig:wTF-muSR}(b) show the 
temperature dependences of $A_{\rm TF}$ and
$\lambda_{\rm TF}$ for 
Na$_{0.75}$CoO$_2$ and Na$_{0.65}$CoO$_2$. Below 
300~K, $A_{\rm TF}$ for Na$_{0.75}$CoO$_2$ is almost
constant ($\sim$ 0.21) 
down to 50~K, then $A_{\rm TF}$ decreases further at
lower $T$, and $A_{\rm TF}$ = 0.018 at 2.5~K, 
while $A_{\rm TF}$ for Na$_{0.65}$CoO$_2$ is 
nearly independent of $T$ down to 2.5~K. On the other
hand, $\lambda_{\rm TF}$ for both samples increases 
slightly with decreasing $T$ due to the 
effect of the nuclear magnetic moments in the
paramagnetic state. 
A marked decrease in $\lambda_{\rm TF}$ is observed
below 50~K only for Na$_{0.75}$CoO$_2$. These results
clearly
indicate that Na$_{0.75}$CoO$_2$ undergoes a magnetic
transition below 50~K. 
Since $A_{\rm TF}$ is roughly proportional to the
volume of paramagnetic 
phases in the sample, the volume fraction $V_{\rm F}$
of the magnetic phase 
at the lowest temperature measured is estimated to be
$\sim$14\%. 

\begin{figure} 
\includegraphics[width=8cm]{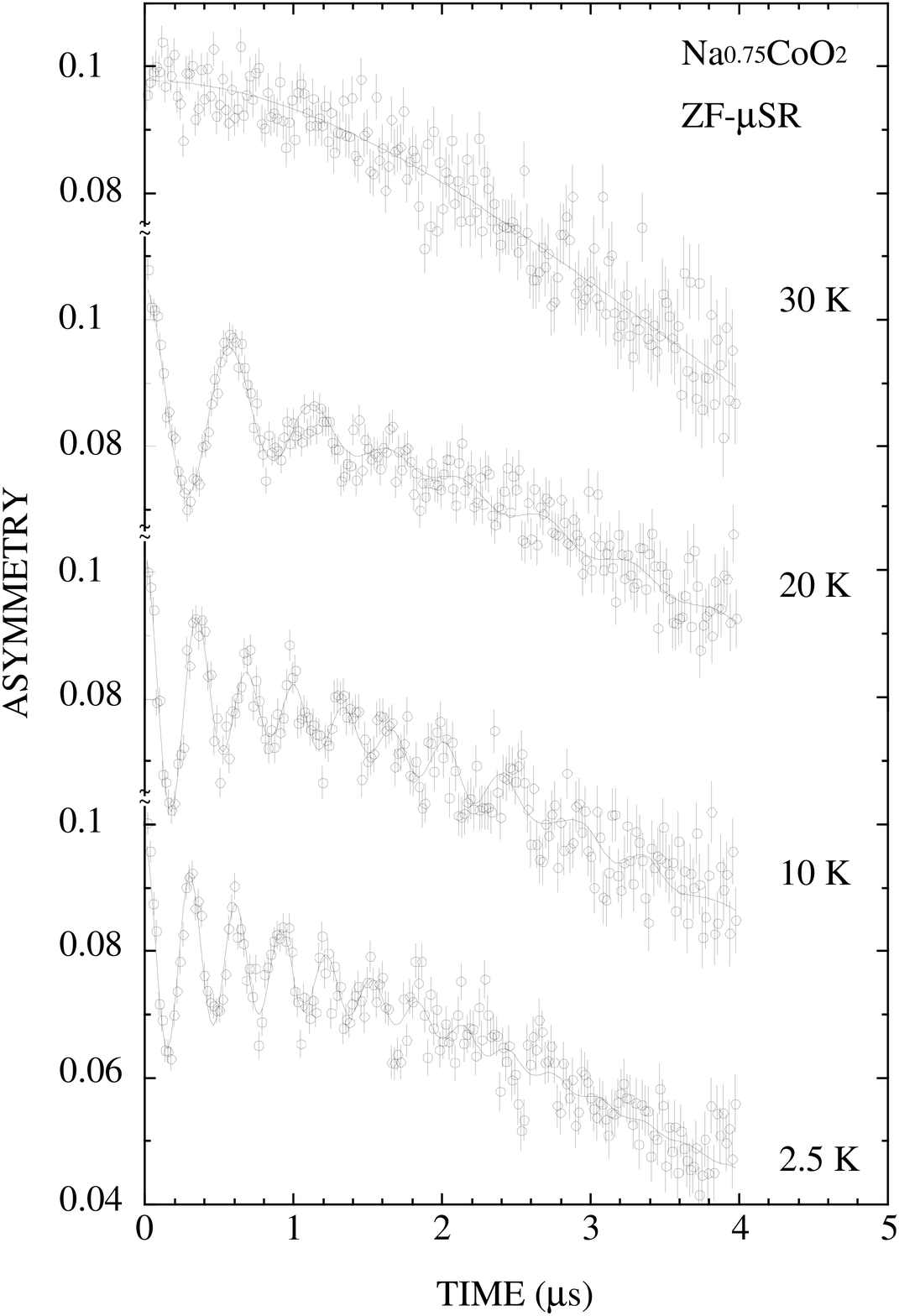}
\caption{\label{fig:ZFmuSRspectra} ZF-$\mu^+$SR time
spectra of
Na$_{0.75}$CoO$_2$ obtained at 30, 20, 10 and 2.5~K;
the solid lines represent 
  the results of fitting using Eq.~(\ref{eq:ZFfit}).} 
\end{figure} 
In order to investigate the magnetism 
in Na$_{0.75}$CoO$_2$ below 22~K in greater detail,
ZF-$\mu^+$SR 
measurements were carried out at 30, 25, 22, 15, 10
and 2.5~K. 
The resulting time spectra, displayed in
Fig.~\ref{fig:ZFmuSRspectra}, 
show a clear oscillation due to quasi-static,
microscopic, internal fields 
at temperatures below $T_{\rm m}$. 

\begin{figure} 
\includegraphics[width=8cm]{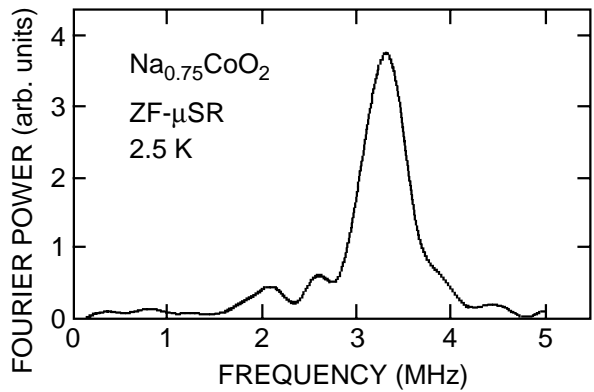}
\caption{\label{fig:FFT} Fourier transform of the
ZF-$\mu^+$SR time 
spectrum from Fig.~\ref{fig:ZFmuSRspectra} at 2.5~K. 
} 
\end{figure} 

Figure~\ref{fig:FFT} shows the Fourier transform of
the ZF-$\mu^+$SR time 
spectrum at 2.5~K. 
There are one main peak at 3.3~MHz and two minor peaks
at 2.6 and 2.1~MHz; 
and the main peak corresponds to the oscillation in 
Fig.~\ref{fig:ZFmuSRspectra}. 

Indeed, the ZF-$\mu^+$SR time spectra were best fitted
with a combination of 
three exponential relaxation functions (for the
signals due to a clear static magnetic field) and 
a Gaussian Kubo-Toyabe function (for the signal from 
muon sites experiencing disordered magnetic fields): 
\begin{eqnarray} 
  A_0 \, P(t) &=& 
    A_1 \, e^{- \lambda_1 t} \, \cos (\omega_{\mu,1} t
+ \phi) \ 
\cr 
  &+& A_{_{\rm KT}} G_{zz}^{\rm KT}(t,\Delta) 
\cr 
  &+&    A_2 \, e^{- \lambda_2 t} \, \cos
(\omega_{\mu,2} t + \phi) \ 
\cr 
  &+&    A_3 \, e^{- \lambda_3 t} \, \cos
(\omega_{\mu,3} t + \phi) \ 
\; , 
\label{eq:ZFfit} 
\end{eqnarray} 
where $A_0$ is the empirical maximum experimental muon
decay asymmetry, 
$A_i$ and $\lambda_i$ ($i$ = 1, 2 and 3) are the 
asymmetries and exponential relaxation rates
associated with the three oscillating signals, 
$A_{\rm KT}$ is the asymmetry of the Gaussian
Kubo-Toyabe signal 
and $\Delta$ is the static width 
of the local frequencies at the disordered sites, and 
\begin{equation} 
  \omega_{\mu,i} \equiv  2 \pi \nu_{\mu,i} =
\gamma_{\mu} \; H_{{\rm int,}i} 
\label{eq:omg} 
\end{equation} 
(where $\gamma_{\mu}$ is muon gyromagnetic ratio) 
is the muon precession frequency in the characteristic
local magnetic field $H_{{\rm int,}i}$ due to the
static magnetic field. 

The static Gaussian Kubo-Toyabe function is 
\begin{eqnarray} 
  &~& G_{zz}^{\rm KT}(t,\Delta) \; = \; {1\over3} 
\cr 
  &+& {2\over3} \, \left( 1 - \Delta^2 t^2 \right) 
  e^{- \Delta^2 t^2 / 2} 
\label{eq:GKT} 
\end{eqnarray} 
 
\begin{figure} 
\includegraphics[width=8cm]{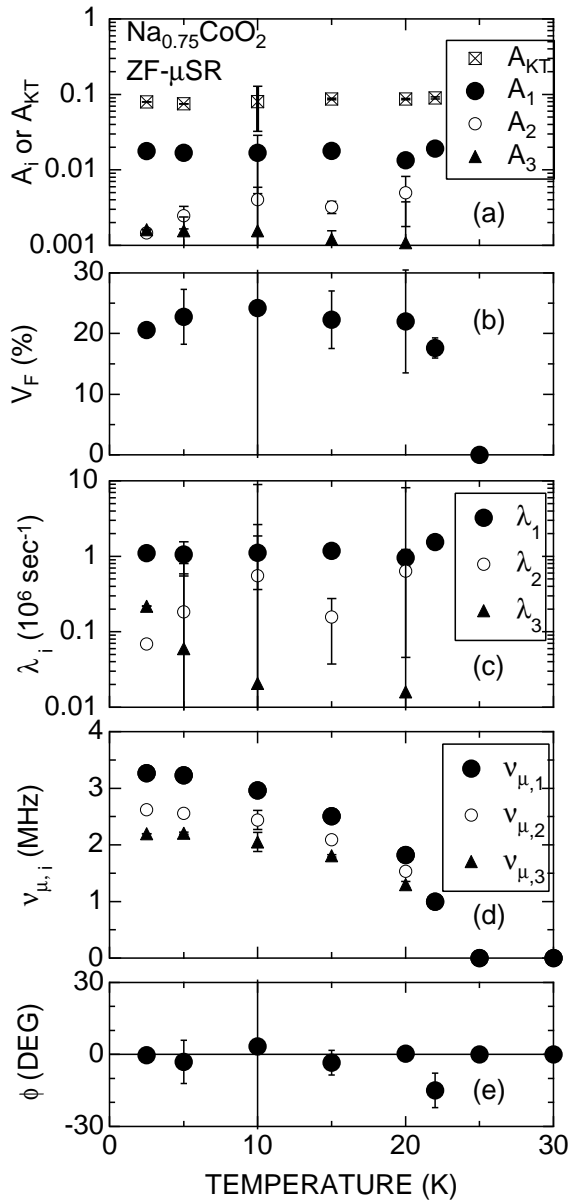}
\caption{\label{fig:ZFmuSR} Temperature dependences of
(a) $A_i$ and $A_{\rm KT}$, 
(b) $V_{\rm F}$, 
(c) $\lambda_i$, 
(d) $\nu_i$ and 
(e) $\phi$ for the Na$_{0.75}$CoO$_2$ sample. 
The data were obtained by fitting the ZF-$\mu^+$SR
time spectra to
Eq.~(\ref{eq:ZFfit}). 
} 
\end{figure}

Figures~\ref{fig:ZFmuSR}(a)-\ref{fig:ZFmuSR}(e) show
the temperature 
dependences of (a) $A_i$ and $A_{\rm KT}$, (b) the
volume fraction of the three exponential relaxation
signals ($V_{\rm F}$), (c) $\lambda_i$, (d) $\nu_i$ and 
(e) $\phi$ in Na$_{0.75}$CoO$_2$. The volume fraction,
$V_{\rm F}$ was calculated as; 
\begin{equation} 
  V_{\rm F} = \frac{\displaystyle
\sum^n_{i=1}A_i}{\displaystyle 
\sum^n_{i=1}A_i+A_{\rm KT}} 
\label{eq:volume} 
\end{equation} 
Below 22~K, $A_i$ and $A_{\rm KT}$ have finite values
and both $A_1$ (main 
component) and $A_{\rm KT}$ are almost constant at
temperatures below 
$T_{\rm m}$. The magnitude of $A_1$ is larger than
those of $A_2$ and $A_3$ 
by one order of magnitude, as expected from the
Fourier transform spectrum 
(see Fig.~\ref{fig:FFT}). However, the ratio between
$A_{\rm KT}$ and $A_1$ 
is $\sim$ 4.5 at 2.5~K; this indicates that $\sim$
80\% muons in the sample 
experience a disordered magnetic field. 
Since the magnetic properties of 
Na$_x$CoO$_2$ are highly sensitive to $x$, 
a slight reduction in $x$ 
makes the sample nonmagnetic in the whole range of
temperature.\cite{12}
Indeed, the wTF-$\mu^+$SR  
experiment on the Na-poor sample, Na$_{0.65}$CoO$_2$, exhibited no
magnetic ordering down to 2.5~K (see Fig.~\ref{fig:wTF-muSR}).
Thus, the Na$_{0.75}$CoO$_2$
sample, although structurally single phase, is found
to be partially nonmagnetic, {\sl i.e.}, magnetically inhomogeneous, 
probably because of a low local Na concentration.
 
The $V_{\rm F}$-{\sl vs.}-$T$ curve is also fairly
flat  
($\sim$ 20\%) below $T_{\rm m}$, similarly to the
$A_1$-{\sl vs.}-$T$ curve. 
This indicates that $\sim$ 20\% of the sample exhibits
the transition to the 
ordered phase at 22~K, and the volume of the ordered
phase does not change down to 
2.5~K. Since $V_{\rm F}$ = 21\% at 2.5~K, it is
concluded that this 
transition is not induced by impurity phases but is an
intrinsic 
behavior of the Na$_{0.75}$CoO$_2$ sample. Moreover,
this suggests that the 
ordered phase is responsible for the change in the
transport properties 
below $T_{\rm m}$. 

The internal magnetic fields of the three signals,
{\sl i.e.}, $\nu_{\mu, 
i}$ with $i$ = 1, 2 and 3, exhibit a similar
temperature dependence. That 
is, as $T$ decreases, each $\nu_{\mu, i}$ increases,
with a decreasing slope d$\nu_{\mu, i}$/d$T$, and
level off to a constant value 
below 5~K. Here, it is worth noting that the
$\nu_{\mu, i}$-{\sl vs.}-$T$ 
curve indicates the change in an order parameter of
the transition. Thus, 
the moderate temperature dependence of $\nu_{\mu, i}$ just below $T_{\rm m}$ 
suggests that the transition is likely to be discontinuous, 
whereas $C_{\rm p}(T)$ supported a continuous transition.\cite{12} 

The values of $\phi$ range between 3.5 and
-16 degrees (see 
Fig.~\ref{fig:ZFmuSR}(e)). This fact, {\sl i.e.},
$\phi \sim 0$, indicates that 
the ordered phase is a either a usual ferromagnet, an
antiferromagnet, a ferrimagnet 
or a commensurate ({\sf C}) spin density wave ({\sf
SDW}) state but not an 
incommensurate ({\sf IC}) {\sf SDW} state, as found
for example for Ca$_3$Co$_4$O$_9$\cite{15} and the Zn and Si doped
CuGeO$_3$ system.\cite{16}
\section{\label{sec:level1}Discussion}

The three exponential relaxation signals and their
$\nu_{\mu, i}$-{\sl 
vs.}-$T$ curves suggest that there are three
inequivalent microscopically ordered muon$^+$
sites in 
the Na$_{0.75}$CoO$_2$ sample. The possible muon$^+$
sites are bound to the oxygen ions in the CoO$_2$ sheets
(the O site) and the two vacant 
sites in the Na planes, namely the Na(1) and Na(2)
sites (see Table 
I).\cite{17} The bond length $d$ of Co-O, Co-Na(1) and
Co-Na(2) are 0.1914, 
0.2703, and 0.3161~nm, respectively. Since the dipolar
field is proportional 
to $d^{-3}$, $H_{\rm int}^{\rm Na(1)}$ and $H_{\rm
int}^{\rm Na(2)}$ should 
be rather small compared with $H_{\rm int}^{\rm O}$.
This is inconsistent 
with the experimental result; that is,  
($d_{\sf Co-Na(2)}/d_{\rm Co-O})^3 \sim 0.22$, 
while $\nu_{\mu,3}$/$\nu_{\mu,1} \sim 0.67$ at 2.5~K
(see Fig.~\ref{fig:ZFmuSR}(d)). 
Thus, muons$^+$ are unlikely to be located at the vacant Na
sites but near the O site. 
The width of the FFT and the relaxation of the oscillating
signals suggest that the field at the O site is 
inhomogeneously broadened probably due to 
variations in the bond length $d_{\rm Co-O}$
caused by the excess Na in the 
Na planes.

\begin{table}
\caption{\label{tab:table1}Structural parameters for Na$_{0.74}$CoO$_2$ with $a$ = 0.2840~nm and $c$ = 1.0811~nm.\cite{17} $g$ means the occupancy of the site.
}
\begin{ruledtabular}
\begin{tabular}{cccccc}
Atom & Site & $g$ & $x$ & $y$ & $z$ \\
\hline
 Na(1) & 2$b$ & 0.23 & 0 & 0 & 1/4 \\
 Na(2) & 2$d$ & 0.51 & 2/3 & 1/3 & 1/4 \\
 Co & 2$a$ & 1.0 & 0 & 0 & 1/2 \\
 O & 4$f$ & 1.0 & 1/3 & 2/3 & 0.0913 \\
\end{tabular}
\end{ruledtabular}
\end{table}

The transition is obviously induced by the ordering of
the Co spins in the 
CoO$_2$ sheets. 
If we assume that the muons experiencing
the ordered field are bound to oxygen, 
then we can estimate the ordered Co
moment as $\sim$ 0.18~$\mu_{\rm B}$ at 
2.5~K,
using $\nu_{\mu, 1}$ = 3.3~MHz and
$d_{\rm Co-O}$ = 
0.1914~nm. 
Considering the number of the nearest
neighboring Co ions for the O 
site (= 3) and the small volume fraction of the
magnetic phase ($\sim$ 
21\%), this value is still 100 times larger than that
estimated by the 
magnetization measurement (1.2 $\times$
10$^{-4}$~$\mu_{\rm B}$ at 
2~K).\cite{12} Such large discrepancy is difficult to
explain based only on 
the present results. 

The related compound, Ca$_3$Co$_4$O$_9$, {\sl i.e.}, 
[Ca$_2$CoO$_3$]$_{0.62}^{\rm RS}$[CoO$_2$] where RS
denotes a rocksalt-type 
subsystem, exhibits two magnetic transitions below
300~K;\cite{15} one is a 
transition to an {\sf IC-SDW} state at $\sim$ 30~K and
the other to a 
ferrimagnetic state at 19~K. The {\sf IC-SDW} is
considered to be induced 
by ordering of the Co moments in the [CoO$_2$]
subsystem, whereas the 
ferrimagnetic ordering is reported to be caused by the
interlayer coupling 
between the Co moments in the [CoO$_2$] and
[Ca$_2$CoO$_3$] 
subsystems\cite{18,19}. 

Therefore, there is a possibility that
Na$_{0.75}$CoO$_2$ below $T_{\rm m}$ 
enters either a ferrimagnet or a commensurate {\sf
SDW} state, because such 
magnetic ordering would decrease the bulk
magnetization drastically. In the 
former case, the Co moments are likely to align
ferromagnetically in the 
CoO$_2$ sheets but antiferromagnetically along the
$c$ axis. In order to 
investigate the magnetism of Na$_x$CoO$_2$ in further
detail, not only 
$\mu^+$SR but also neutron diffraction and
$^{59}$Co-NMR measurements are 
necessary for single crystals with various $x$. 

\section{\label{sec:level1}Summary}

We measured positive muon-spin-rotation and relaxation
($\mu^+$SR) spectra 
in a polycrystalline Na$_{0.75}$CoO$_2$ sample below
300~K. At 
temperatures below 22~K (= $T_{\rm m}$), zero field
$\mu^+$SR spectra 
exhibited clear oscillations due to static internal
magnetic fields, 
although the volume fraction of the magnetically
ordered phase was only $\sim$21\% at 2.5~K. 
Furthermore, the Co moment estimated by the present
$\mu^+$SR experiment 
was $\sim$ 100 times larger than that  estimated from
the magnetization measurement. 
This suggested that the ordered phase is in either a
ferrimagnet or a 
commensurate spin density wave state. 
In addition, a large fraction of the muons, given
by $A_{\rm KT}$  (KT background) was found to
experience a broad distribution of fields, perhaps
reflecting the disorder due to the excess
Na.

\begin{acknowledgments} 
We would like to thank Dr. S.R. Kreitzman, Dr. B. Hitti
and Dr. D.J. Arseneau 
of TRIUMF for their help with the $\mu^+$SR experiment. 
Also, we appreciate Dr. T. Tani and Dr. R. Asahi of
Toyota Central R\&D Labs., Inc. 
for fruitful discussions. 
This work was partially supported by 
the Canadian Institute for Advanced Research, 
the Natural Sciences and Engineering Research Council of Canada 
and (through TRIUMF) the National Research Council of Canada. 
\end{acknowledgments}


\end{document}